\documentclass[12pt]{iopart}

\usepackage[dvips]{graphicx}
\usepackage{color}
\usepackage{ulem}
\def\be{\begin{equation}}
\def\ee{\end{equation}}
\def\bea{\begin{eqnarray}}
\def\eea{\end{eqnarray}}
\def\kv{{\bf k}}

\def\Sv{{\bf S}}
\def\Mv{{\bf M}}

\def\sv{{\bf s}}
\def\Rv{{\bf R}}
\def\rv{{\bf r}}
\def\sigmav{{\vec\sigma}}

\def\be{\begin{equation}}
\def\ee{\end{equation}}

\begin{document}
\title{Anisotropic conductivity in magnetic topological insulators}

\author{A. Sabzalipour, J. Abouie, and S. H. Abedinpour}
\address{Department of Physics, Institute for Advanced Studies in Basic Sciences (IASBS), Zanjan 45137-66731, Iran}
\begin{abstract}
We study the surface conductivity of a three dimensional topological insulator doped with magnetic impurities. The spin-momentum locking of surface electrons makes their scattering from magnetic impurities anisotropic and the standard relaxation time approximation is not applicable. Using the semiclassical Boltzmann approach together with a generalized relaxation time scheme, we obtain closed forms for the relaxation times and analytic expressions for the surface conductivities of the system as functions of the bulk magnetization and the orientation of the aligned surface magnetic impurities.
We show that the surface conductivity is anisotropic, and strongly depends both on the direction of the spins of magnetic impurities and on the magnitude of the bulk magnetization. In particular, we find that the surface conductivity has its minimum value when the 
spin of surface impurities are aligned perpendicular to the surface of TI, and therefore the backscattering probability is enhanced due to the magnetic torque exerted by impurities on the surface electrons.
\end{abstract}
\maketitle

\section{Introduction}\label{sec:intro}

Topological insulators have been fascinating both from the fundamental physical and applied technological points of view.
While the bulk of a D-dimensional topological insulator (TI), almost similar to any other band insulator is gapped, (D-1)-dimensional gapless states emerge on its surfaces or edges~\cite{TI_book, hasan_rmp, zhang_rmp}. 
These lower dimensional metallic surface or edge states are protected via a particular, \textit{e.g.,} time reversal (TR) or crystalline~\cite{fu_tci} symmetries. 
A large family of topological insulators, appearing in materials with large spin-orbit couplings are time-reversal invariant three dimensional insulators (3D-TI). 
It turns out that the surface states of these materials at low energies, could be readily described with an effective 2D massless Dirac Hamiltonian~\cite{TI_book}. 
The surface of some 
compounds like $\mathrm{Bi_2 Se_3}$ and $\mathrm{Bi_2 Te_3}$, in particular have simpler structures, and support a single Dirac-cone at their $\Gamma$-points~\cite{xia_natphys_2009, zhang_natphys_2009}.
At higher energies, however, corrections like quadratic-in-momentum  and anisotropic hexagonal-warping effects should be included in the effective Hamiltonian~\cite{Alpich_prl}.
The chiral nature of these Dirac states have some peculiar implications on their transport properties~\cite{culcer_prb_2010,culcer_review}. The back-scattering of these surface electrons from impurities, and therefore the localization is forbidden as long as the impurity potential does not break the TR symmetry~\cite{roushan_nat_2009}.

On the other hand, the story is totally different if a TI is doped with magnetic impurities. 
Magnetic impurities break the TR invariance and the back-scattering is not prohibited anymore. The amplitude of the back-scattering could be controlled by the orientation of the magnetic moment of the surface impurities and, by the magnitude of the mass gap induced by the net magnetization of system. Consequently the surface conductivity would become anisotropic, and strongly depends on the magnitude and orientation of the magnetization. 
Bulk conductivity, on the other hand, is not necessarily sensitive to these details of the magnetic impurities and therefore magneto-transport measurements are very likely to be able to distinguish between the bulk and surface contributions to the conductivity of 3D topological insulators. 

In this paper, we study the surface charge conductivity of a 3D-TI doped with magnetic impurities using a semiclassical Boltzmann approach.
The spin-momentum locking of the surface electrons makes their scattering from magnetic impurities anisotropic and the standard relaxation time approximation (RTA) is not applicable.  Many attempts have been devoted to the development of a generalized RTA for anisotropic systems~\cite{schliemann_prb_2003,rushforth_prl_2007,trushin_prb_2007,Jairo}.
However, it is still felt the lack of a closed form for the relaxation times of 
topological insulators with anisotropic scatterings. 
In this paper, following the general recipe of 
Ref.~\cite{Jairo}, we present a closed form for the relaxation times of surface electrons of magnetic topological insulators, 
obtain the surface charge conductivity of the system, and show that it is anisotropic and strongly depends both on the direction of the spins of magnetic impurities and on the magnitude of the bulk magnetization. 
In particular, we find that the surface conductivity has its minimum value when the surface impurities' spins are aligned perpendicular to the surface of TI, and therefore the back-scattering probability is enhanced due to the magnetic torque exerted by impurities' spins on the spin of surface electrons. 
We also observe an interesting effect when the spin of surface impurities is parallel to the surface.
In this case the surface conductivities become independent of the magnitude of the surface energy gap.

We have organized the rest of this paper as follows.
In Sec.~\ref{sec:model}, we first introduce the effective model 
of a three-dimensional topological insulator in the presence of magnetic impurities, and then explain the Boltzmann formalism and the generalized relaxation time approximation we use to study the anisotropic conductivities. Our results for the charge conductivity of massless and massive surface carriers are discussed in Sec.~\ref{sec:results}. In Sec.~\ref{sec:summary} we summaries and conclude our main findings. Finally, the details of obtaining the generalized relaxation times are presented in an Appendix.

\begin{figure}[t]
\centering
\includegraphics[width=100mm]{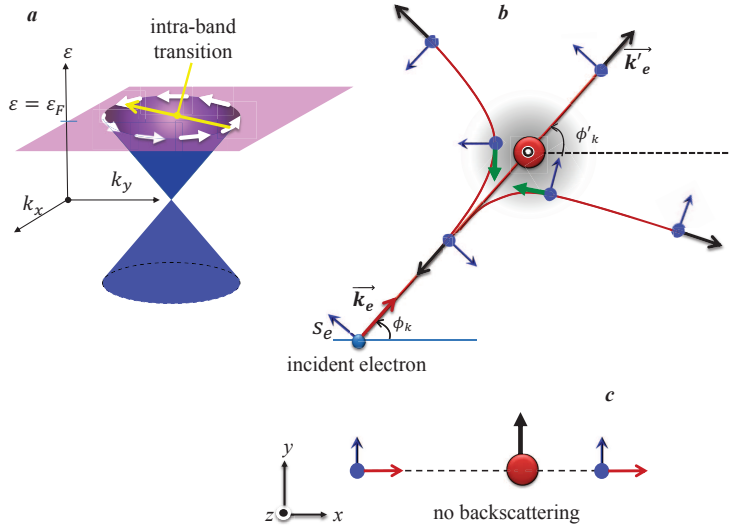}
\caption{(Color online) (a) Band structure of gapless Dirac electrons. The yellow arrow shows the intra-band scattering from an initial state ${\bf k}$ to a final state ${\bf k}'$ on the Fermi surface. (b) and (c) are the schematic plots of the scattering of a surface electron from magnetic impurities in which their spins are aligned in the $z$ and $y$ directions, respectively. The blue solid spheres are the incident and scattered electrons, the red and blue arrows are the momentum and spin of itinerant electrons, respectively. The green arrows show the directions of exerted torque ($\Sv\times\sv$) on the surface electrons. In panel (b) the spin of magnetic impurity is along the $z$ axis and the backscattering process is more probable, while in (c) the torque is zero and therefore the backscattering is forbidden. In this case the conductivities have their maximum values.
\label{schematic}}
\end{figure}

\section{Model and Formalism}\label{sec:model}

The \textit{minimal} effective Hamiltonian describing the surface electrons of a three-dimensional topological insulator is given by;
\begin{equation} \label{H1}
{\cal H}_{\rm D}=\hbar v_{\rm F} \sum_{\kv} \psi^\dagger_\kv (\kv \times {\vec\sigma})_z \psi_\kv ~,
\end{equation}
where the $z$-direction is chosen normal to the surface of topological insulator, $\psi^\dagger_\kv$ and $\psi_\kv$ are two-component creation and annihilation operators of electrons with wave vector $\kv=(k_x,k_y)$, $v_{\rm F}$ denotes the Fermi velocity of surface electrons, and $\sigmav=(\sigma_x,\sigma_y,\sigma_z)$ is the vector of Pauli matrices for itinerant electrons.
The above Hamiltonian is indeed the two-dimensional massless Dirac Hamiltonian, which arises from the spin-orbit coupling of surface electrons and causes the helical order of surface states.
In writing the effective Hamiltonian~(\ref{H1}), the quadratic and \textit{anisotropic} cubic terms in momentum have been omitted, which is well justified in the low carrier concentrations~\cite{TI_book, hasan_rmp, zhang_rmp}.

When the topological insulator is doped with magnetic impurities, we assume that the exchange interaction between surface electrons and magnetic moments are \textit{isotropic} and can be written as
\be\label{h_ex}
{\cal H}_{{\rm ex}}=\sum_{\Rv,\rv}J(\Rv-\rv)~\Sv(\Rv) \cdot \sv(\rv)~,
\ee
where  $\Sv(\Rv)$ is the spin of the magnetic impurity located at $\Rv$, $\sv(\rv)=\psi^\dagger(\rv) \sigmav \psi(\rv)$ is the spin of the surface electron at $\rv$, with $\psi^\dagger(\rv)$ and $\psi(\rv)$ being the creation and destruction field operators, and $J(r)$ is the exchange coupling parameter.
Summations in Eq.~(\ref{h_ex}) run over the positions of magnetic impurities $\Rv$ and surface electrons $\rv$.

Naturally, magnetic impurities got placed both on the surface, and in the bulk of TI.
The exchange interaction of surface electrons with bulk magnetic impurities aligns the spin of impurities which produces a magnetization in the bulk of the system. This bulk magnetization decreases by increasing the temperature and falls to zero at a bulk critical temperature $T^{bulk}_c$.
The value of $T^{bulk}_c$ generally depends on the details of band structure and other factors.
For ${\rm Bi_2Se_3}$ with $5\%$ concentration of ${\rm Cr}$ dopants, Yu \textit{et al.},~\cite{Yu} estimate $T_c^{bulk} \sim 70 K$ using first-principles numerical calculations.
A more recent and comprehensive study of magnetically doped ${\rm Bi_2 Se_3}$, ${\rm Bi_2 Te_3}$, and ${\rm Sb_2 Te_3}$ using first-principles Green's function method predicts even higher transition temperatures for higher concentrations of the magnetic dopings~\cite{vergniory_prb_2014}.
The bulk magnetization generates a temperature dependent gap on the spectrum of surface states which is proportional to the bulk magnetization.
In order to capture these effects qualitatively, we will separate the contribution of bulk and surface impurities in the exchange Hamiltonian~(\ref{h_ex}), while applying the mean-field approximation (MFA) to the bulk impurity part of it, we will keep surface impurities as the source of scattering for surface electrons, which later will be treated within the Boltzmann formalism.
However, note that this separation is not necessary and both bulk and surface magnetic impurities can in principle induce a finite gap on the surface states~\cite{henk_prl_2012}, but in the following we will treat bulk and surface impurities separately only for the sake of definiteness.

Now, the interaction of surface electrons with bulk impurities within the MFA reads
\bea
\nonumber{\cal H}_{\rm M} &=&\sum_{\rv,\Rv_b}J(\Rv-\rv)\Sv(\Rv) \cdot \sv(\rv)
\stackrel{\rm MFA}{\approx} \Mv \cdot \sigmav~,
\eea
where ${\bf R}_b$ shows the position of magnetic impurities located in the bulk of TI. The bulk magnetization $\Mv$ could be determined self-consistently~\cite{rosenberg_prb_2012}.
As for the contribution from the surface impurities, furthermore assuming that the exchange coupling between surface impurities and electrons is short range, \textit{i.e.}, $J(\Rv-\rv) \approx J\delta(\Rv-\rv)$, one finds
\be\label{eq:v_scatt}
V_{\rm scat} \approx \frac{J}{A} \sum_j \Sv_j \cdot \sv_j~,
\ee
where $A$ is the TI's surface area and the summation runs over the position of surface impurities.
Also note that as Kondo like effects~\cite{zare_prb_2013} are beyond the scope of current paper, we will treat the spins of surface impurities in Eq.~(\ref{eq:v_scatt}) classically.
Now, the total Hamiltonian of surface electrons in the presence of magnetic impurities can be written as ${\cal H}={\cal H}_0+V_{\rm scat}$, 
where ${\cal H}_0={\cal H}_{\rm D}+{\cal H}_{\rm M}$.
The $z$-component of the bulk magnetization $\Mv$ breaks the time reversal symmetry of the surface states and opens an energy gap at the Dirac point.
For the special case of $\Mv=M \hat{e}_z$, the eigenvalues and eigenvectors of ${\cal H}_0$ are
\be
\varepsilon^{\rm M}_\pm(k)=\pm\varepsilon^{\rm M}_k= \pm \sqrt{(\hbar v_{\rm F} k)^2+M^2}~,
\ee
and
\be
\psi^{\rm M}_{\kv,\pm}(\rv)=\frac{e^{i \kv\cdot \rv}}{\sqrt{A(1+\xi_\kv^{\pm 2})}}
\left(
\begin{array}{c}
e^{-i \phi_\kv/2} \\
\pm i \xi_k^{\pm 1} e^{i \phi_\kv/2}
\end{array}
\right)~,
\ee
where $\xi_k=\sqrt{(\varepsilon_k^{\rm M}-M)/(\varepsilon_k^{\rm M}+M)}$ 
and $\phi_\kv=\arctan({k_y}/{k_x})$ shows the direction of the wave vector of surface electrons.
The in-plane component of the magnetization, on the other hand, simply shifts the position of the Dirac point in the $(k_x,k_y)$-plane, \textit{i.e.,} is a pure gauge, and hence has no impact on the physical observables of the system.
Therefore wherever we talk about a non-zero bulk magnetization, we will simply  consider its component perpendicular to the surface.
The magnetization and thus the energy gap strongly depend on thermal fluctuations in the system and vary with temperature.
The massive Dirac electrons with variant mass causes the emergence of different exotic phenomena. In the following we study these effects on the transport properties of the system.

\subsection{Generalized relaxation times and Anisotropic conductivity}

In order to investigate the behavior of surface electrons, which weakly interact with dilute surface magnetic impurities, we model the surface electrons with ${\cal H}_0$ and study their scattering from surface impurities using the semiclassical Boltzmann formalism.
We write the equation for the non-equilibrium distribution function $f$ of itinerant electrons, in the presence of
a uniform external electric field $\mathbf{E}$ as~\cite{mahan_nutshell}
\begin{equation}\label{b1}
\left(\frac{\partial f}{\partial t}\right)_\mathrm{scat}=
|e|\mathbf{E}\cdot\mathbf{v}_{\kv}
\left(-\frac{\partial f^0}{\partial\varepsilon_{\kv}}\right)~,
\end{equation}
where $\mathbf{v}_{\kv}$ and $\varepsilon_{\kv}$ are the velocity and energy of the incident wave packet with wave vector $\kv$, respectively and the equilibrium distribution $f^0(\varepsilon_{\kv})$ is the Fermi-Dirac function.
Considering only elastic scatterings,
which in our single and isotropic band regime implies $|\kv|=|\kv^{\prime}|$, and using the detailed balance, we find
\begin{equation}\label{b2}
\left(\frac{\partial f}{\partial t}\right)_{\rm scat}
=\sum_{\kv'}   W_{\kv,\kv^{\prime}}\left[f(\varepsilon_{\kv^{\prime}})- f(\varepsilon_{\kv})\right]~,
\end{equation}
where $W_{\kv,\kv^{\prime}}$ is the transition probability between $\kv$ and $\kv^{\prime}$ states.
Using the Fermi's golden rule, $W_{\kv,\kv^{\prime}}$ reads
\begin{equation}\label{b3}
W_{\kv, \kv'}=\frac{2\pi}{\hbar} \left| T_{\kv,\kv'}\right|^2
\delta\left(\varepsilon_{\kv}-\varepsilon_{\kv'}\right)~.
\end{equation}
Here, $T_{\kv,\kv'}$ is the T-matrix and in the regime of dilute doping, one can resort to the first Born approximation
$T_{\kv,\kv'} \approx V_{\kv,\kv'}= \langle\kv|V_{\rm scat}|\kv'\rangle$, where $|\kv\rangle$ and $|\kv'\rangle$ are the eigenstates of the Hamiltonian ${\cal H}_0$.
Throughout this work we will consider uncorrelated random distribution of the impurities~\cite{kohn_luttinger} and will assume that their spin are aligned in the same direction. Therefore the square of T-matrix simplifies to $\left|T_{\kv,\kv'}\right|^2=(n_{\rm imp}/A) J^2 \left|\langle \kv | \Sv \cdot \sigmav | \kv' \rangle\right|^2,$
with $n_{\rm imp}$ being the impurities concentration. 
Note that
 as the spins of impurities are considered classically,
 the $T$-matrix depends only on the square of $J$ and therefore the sign of the coupling parameter $J$, \textit{i.e.,} the ferromagnetically or antiferromagnetically coupling of Dirac electrons with surface impurities has no effect on the transition probability. 
When the scattering potential is isotropic, the transition probability will depend only on the angle between $\kv$ and $\kv'$, and obtaining the transport relaxation time $\tau_\kv$ and the charge conductivity would be 
straightforward~\cite{mahan_nutshell}.
This is indeed the case when \textit{e.g.}, the spin of surface impurities are all aligned perpendicular to the surface \textit{i.e.},
$\Sv=S {\hat e}_z$.
In this case one can continue with the standard relaxation time approximation recipe~\cite{mahan_nutshell}, assuming that the relaxation time depends only on the magnitude of $\kv$, arrives at the renowned expression
$1/{\tau_{\kv}}=\sum_{\kv'} W_{\kv,\kv'}[1-\cos(\phi_{\kv}-\phi_{\kv'})]$.
On the other hand, when the spins of surface impurities are not aligned in the $z$-direction, due to the interaction of helical electrons with the in-plane components of the spin of surface impurities, the scattering potential becomes anisotropic and the transition probability explicitly depends on the direction of both incident and scattered electrons' wave packets. Consequently the relaxation time is strongly anisotropic and depends on the magnitude and direction of $\kv$, and on the orientations of the surface magnetic impurities. As a result, one can not follow the standard relaxation time scheme anymore.
In order to capture the effects of this anisotropy in the charge conductivity and other transport properties of the system, we approximate the non-equilibrium distribution function as
\begin{equation}\label{b4}
\delta f(\phi_\kv, \chi)=e E v_{\kv}\left[\tau_1(\phi_\kv)\cos\chi+\tau_2(\phi_\kv)\sin\chi\right]
\frac{\partial f^0(\varepsilon)}{\partial  \varepsilon}~,
\end{equation}
where $\delta f=f-f^0$ is the deviation of the distribution function from the equilibrium Fermi-Dirac distribution and $\chi$ is the angle of electric field  with the $x$-axis~\cite{Jairo}. The two independent coefficients $\tau_1$ and $\tau_2$ are the generalized relaxation times. 
Now, using Eqs.~(\ref{b1}),~(\ref{b2}), and (\ref{b4}), the relaxation times read
\be\label{rt-aniso}
\left\{
\begin{array}{c}
\tau_1(\phi_{\kv}) {\bar w}(\phi_{\kv}) =\cos\phi_{\kv}+\sum_{\kv'}W(\phi_{\kv},\phi_{\kv'})\tau_1(\phi_{\kv'})\\ \\
\tau_2(\phi_{\kv}) {\bar w}(\phi_{\kv}) =\sin\phi_{\kv}+\sum_{\kv'}W(\phi_{\kv},\phi_{\kv'})\tau_2(\phi_{\kv'})
\end{array}
\right.~,
\ee
where ${\bar w}(\phi_\kv)=\sum_{\kv'} W(\phi_\kv,\phi_{\kv'})$ is the transition probability averaged over all out-going directions.
Note that here, for the sake of brevity, the apparent dependance of the relaxation times and other quantities to the magnitude of wave vectors are dropped. As $|\kv|=|\kv'|$ in the elastic scattering processes, this should not cause any confusion.

The relaxation times could be obtained by replacing $\tau_1$ and $\tau_2$ with their Fourier expansions on both sides of Eq.~(\ref{rt-aniso}). 
Now, substituting the relaxation times in Eq. (\ref{b4}) to find $\delta f$, the conductivity is obtained from
\begin{equation}\label{current}
\sigma_{\alpha\beta}=-\frac{e}{AE_{\beta}}\sum_{n,\kv}v_{n\alpha}(\kv) [f^0(\kv)+\delta f(\kv)]~,
\end{equation}
where $\alpha$ and $\beta=x,y$ and  $v_{n\alpha}(\kv)=\nabla^{\alpha}_{\kv}\varepsilon_{n}(\kv)/\hbar$ is the group velocity of the $n$-th band.
Note that the anomalous contribution to the velocity, arising from the band topology is omitted here, as it is only responsible to an intrinsic contribution to the transverse conductivity, which is independent from the strength and direction of the magnetic impurities.

\section{Results and Discussions}\label{sec:results}

The scattering of a surface electron from the impurities located at the surface of a TI, is different for massless and massive surface electrons. 
Following we will discuss them separately. 
Moreover, without loss of generality, we will assume that the TI is electron doped so the chemical potential $\mu$ is located in the conduction band of surface state ($\mu>0$), but still resides inside the bulk gap.

\begin{figure}
\centering
\includegraphics[width=90mm]{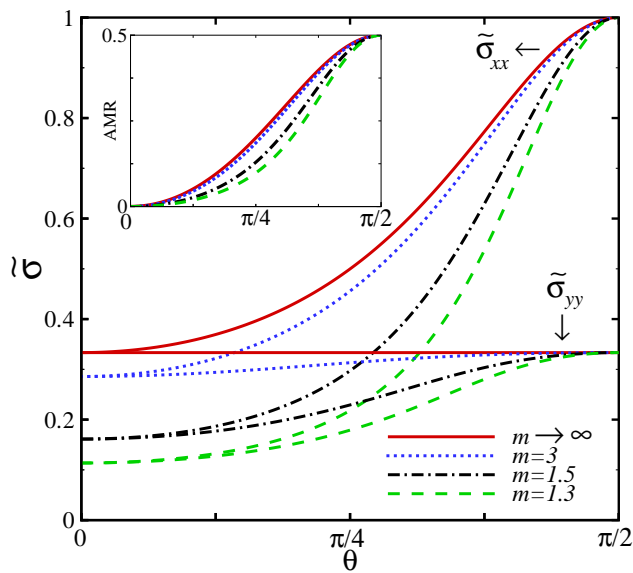}
\caption{(Color online)
The dimensionless surface conductivities (${\tilde \sigma}=\sigma/\sigma_0$) of a magnetic topological insulator in terms of the surface impurities' orientation $\theta$ for different values of the gap parameter $m$. The azimuthal angle of impurities is set to $\phi=\pi/2$. Note that the solid lines (\textit{i.e.}, $m\to \infty$) correspond to a gapless system.
Inset: the anisotropic magneto-resistance of the same system. \label{cond-gapped1}}
\end{figure}

\subsection{Surface conductivity in the absence of bulk magnetization ($M=0$)}

Let us first investigate the surface conductivity of a TI at $T>T_c^{bulk}$, where the bulk magnetization $M$ is zero, and therefore the energy spectrum is gapless.
Because of the elastic scattering of itinerant electrons by localized surface magnetic impurities, the inter-band transition is forbidden.  Then, valance band contribution to the conductivity would be negligible as long as $k_{B}T$ is much less than the chemical potential.
Within the first Born approximation, the square of T-matrix  is given by
\begin{equation}\label{T-gapless}
\left|T_{\kv, \kv'}\right|^2=\frac{n_{\rm imp}J^2S^2}{A}
\left(\sin^2\theta \cos^2 \phi_++ \cos^2\theta \sin^2 \phi_- \right)~,
\end{equation}
where $\phi_\pm=\left(\phi_{\kv'}\pm\phi_\kv\right)/2$ and $\theta$ is the tilting angle of the spin of impurities with respect to the surface normal vector.
Moreover, in arriving at Eq.~(\ref{T-gapless}), we have assumed that the spin of all surface impurities are aligned in the same direction and without lose of generality we have defined our coordinate system such that the magnetic moments lie on the $y-z$ plane (\textit{i.e.}, $S_x=0$).

Now, using Eqs.~(\ref{T-gapless}) and (\ref{b3}) in Eqs.~(\ref{rt-aniso}), the generalized relaxation times read (for details, see, ~\ref{sec:appendix})
\be
\tau_1(\phi_\kv)=\frac{ \tau_\kv^0 \cos\phi_\kv}{2+\cos 2 \theta}~,~~~
\tau_2(\phi_\kv)=\frac{\tau_\kv^0 \sin\phi_\kv}{3}~,
\ee
where $\tau_\kv^0=4\hbar^3 v^2_{\rm{F}}/(n_{\rm{imp}} J^2 S^2 \varepsilon_k)$ has the units of time.
The charge conductivity (in the units of $e^2/h$) at low temperature could be also readily obtained as
\be\label{sigma_gapless}
\sigma_{xx}=\frac{\sigma_0}{2+\cos 2\theta} ~,~~
\sigma_{yy}=\frac{\sigma_0}{3} ~,~~
\sigma_{xy}=\sigma_{yx}=0~,
\ee
with $\sigma_{0}= \frac12(\varepsilon_k \tau_\kv^0/\hbar)=2\hbar^2 v^2_{\rm F}/(n_{\rm imp}J^2S^2)$.
As we had already anticipated, for $S_y=0$ (\textit{i.e.}, $\theta=0$), the T-matrix depends only on the angle between $\kv$ and $\kv'$ and the relaxation time and charge conductivity become isotropic: $\sigma_{xx}=\sigma_{yy}=\sigma_0/3$.
When
the momentum of incident electron is along the y-direction,
the back-scattering (BS) transition probability $W^\textrm{BS}=W_{\kv,-\kv}$ is constant, \textit{i.e.}, $W^\textrm{BS} \propto n_{\rm imp}J^2S^2$, and it is natural to expect that the charge conductivity 
$\sigma_{yy}$
should be independent of the tilting angle, while for the
$x$-direction
$W^\textrm{BS} \propto n_{\rm imp}J^2S^2\cos^2\theta$, and increasing the tilting angle from $0$ to $\pi/2$, the back-scattering transition probability decreases and consequently the conductivity $\sigma_{yy}$ increases from $\sigma_0/3$ to $\sigma_0$.
This could be easily understood, at least qualitatively, in terms of the torque exerted on the spin of itinerant electrons by spin of magnetic impurities. Without magnetic impurities, the back-scattering is essentially prohibited due to the spin-momentum locking of the surface states. The spin of a magnetic impurity, on the other hand, can rotate the spins of electrons, allowing them to backscatter which consequently suppresses the conductivity (see, Fig.~\ref{schematic}).
When our system is gapless, the spins of surface electrons are perpendicular to both their momentum and to the $z$-direction.
When the electric field is applied in the $x$-direction, the main contribution to the charge transport comes from electrons whose momentum are along the $x$-axis. The spins of these electrons are along the $y$-axis and as the exerted magnetic torque is proportional to $(\Sv \times \sv)$, increasing the tilting angle of magnetic impurities' spin from $0$ to $\pi/2$, the magnitude of the torque drops from its maximum value to zero. Therefore the backscattering probability becomes smaller and the charge conductivity enhances.
On the other hand, when the electric field is applied along the $y$-axis, the spins of surface electrons moving along the electric field are always perpendicular to the direction of spin of impurities, making the magnetic torque maximum, independent of the tilting value. So the charge conductivity $\sigma_{yy}$ is also constant.

We have illustrated the scaled conductivity of a topological insulator doped with spin-$S$ magnetic impurities in Fig.~\ref{cond-gapped1} (solid lines).
We have also plotted in the inset of Fig.~\ref{cond-gapped1} the anisotropic magneto-resistance (AMR) versus $\theta$. AMR is indeed an indicator of the anisotropy of the surface conductivity and is defined as~\cite{Jairo}
\begin{equation}\label{AMR-gapless}
AMR=\frac{\sigma_{xx}-\sigma_{yy}}{\sigma_{xx}+\sigma_{yy}}
=\frac{\sin^2\theta}{2+\cos^2\theta}~.
\end{equation}
Since for gapless surface $\sigma_{yy}$ is constant, the behavior of AMR is similar to that of $\sigma_{xx}$ and the anisotropy enhances with increasing the tilting angle. At $\theta=\pi/2$, when the spin of all impurities lie on the surface of the topological insulator, the AMR has its maximum value of $0.5$.

\subsection{Surface conductivity in the presence of a finite bulk magnetization ($M\neq 0$)}

\begin{figure}
\centering
\includegraphics[width=90mm]{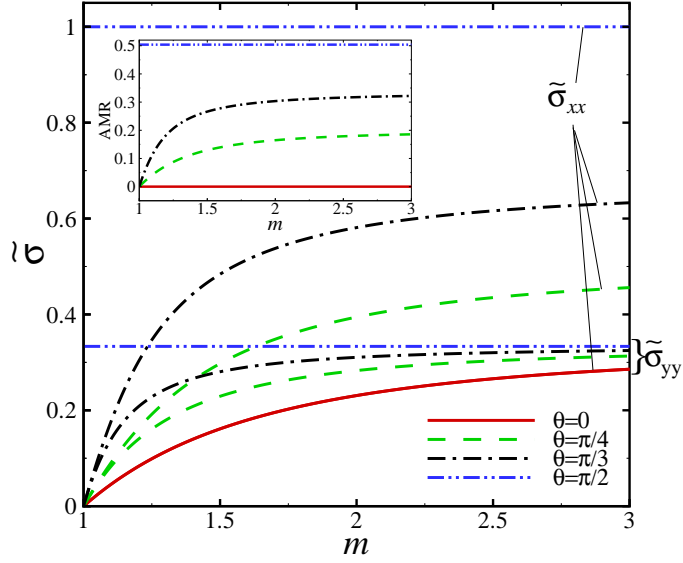}
\caption{(Color online)
The dimensionless surface conductivities (${\tilde \sigma}=\sigma/\sigma_0$) and anisotropic magneto-resistance (the inset) of a magnetic topological insulator versus $m$ for different orientations of the spin of the surface impurities $\theta$. At $\theta=0$, all the impurities' spin are perpendicular to the surface and $\sigma_{xx}$ and $\sigma_{yy}$ coincide (solid red line). \label{cond-gapped2}}
\end{figure}

Decreasing the temperature, a topological phase transition occurs at the critical temperature $T^{bulk}_c$, where the magnetization of the bulk impurities $M$ becomes nonzero.  
This breaks the time reversal symmetry of the system and an energy gap of $\Delta= 2M$ opens up in the surface states~\cite{chen_science_2010}.
This gap increases by lowering the temperature as $|T-T^{bulk}_c|^{-0.5}$ and reaches its saturation value at zero temperature.
While this TR breaking makes the system topologically trivial, but as the net magnetization would be very small for dilute dopings, the surface gap could be essentially much smaller than the bulk one. 
In this regime, still the chiral nature of the surface states would play an important rule in different properties of the system.

Within the first Born approximation, the square of T-matrix of the gapped system reads
\bea\label{T-gapped}
\nonumber\left|T_{\kv,\kv'}\right|^2=&\frac{n_{\rm imp}}{A}J^2 S^2  \left|
(1-\gamma_k^2)^{1/2} \sin \theta \cos\phi_+ \right. \\
&\left.+\gamma_k \cos\theta \cos\phi_- + i \cos \theta \sin\phi_- \right|^2~,
\eea
where $\gamma_k=M/\epsilon_k^M$.
 Note that for $M\to 0$, the gapless result of Eq.~(\ref{T-gapless}) is reproduced. 
Now, we can find $W_{\kv,\kv'}$ from Eq.~(\ref{b3}), and then the relaxation times from Eqs.~(\ref{rt-aniso}), which their general closed forms could be expressed as
\bea
\nonumber&&\tau_1(\phi_{\kv})=\frac{1}{{\bar w}(\phi_\kv)}\left[1+\frac{w^c/\cos\phi_\kv+(\gamma_k^2-1)
\cos2\theta}{1+\Gamma+(1+\Gamma\gamma_k^2)\cos2\theta}\right]\cos\phi_\kv~,\\
&&\tau_2(\phi_{\kv}) =\frac{1}{{\bar w}(\phi_\kv)}\left[1+\frac{\gamma_k^2-1}{1-\gamma_k^2+w^0(1+\Gamma)}\right]\sin\phi_\kv~,
\eea
where $w^0$, $w^c$, $\Gamma$ and ${\bar w}(\phi_\kv)$ are defined in \ref{sec:appendix}. 

Finally, the conductivities at zero temperature are readily obtained
\bea
\nonumber\sigma_{xx}&=\frac{2(m^2-1) }{4m^2\cos^2\theta+g[m^2+\cos2\theta]}\sigma_0~, \\
\nonumber\sigma_{yy}&=
\frac{(m^2+\cos2\theta)[g-2]}{\cos4\theta-1+(m^2+\cos2\theta)[g-2]}\sigma_0~,\\
\sigma_{xy}&=\sigma_{yx}=0~,
\eea
where $m=\mu/M$ is defined as the inverse of the Dirac electron mass and
\be
g=\frac{\sqrt{4(m^4+1)+2m^2\left(4\cos2\theta+\cos4\theta-1\right)}}{\left|m^2+\cos2\theta\right|}~.
\ee
In Figs.~\ref{cond-gapped1} and \ref{cond-gapped2} we have plotted the surface conductivities versus $\theta$ and $m$, respectively.

\begin{figure}
\centering
\includegraphics[width=100mm]{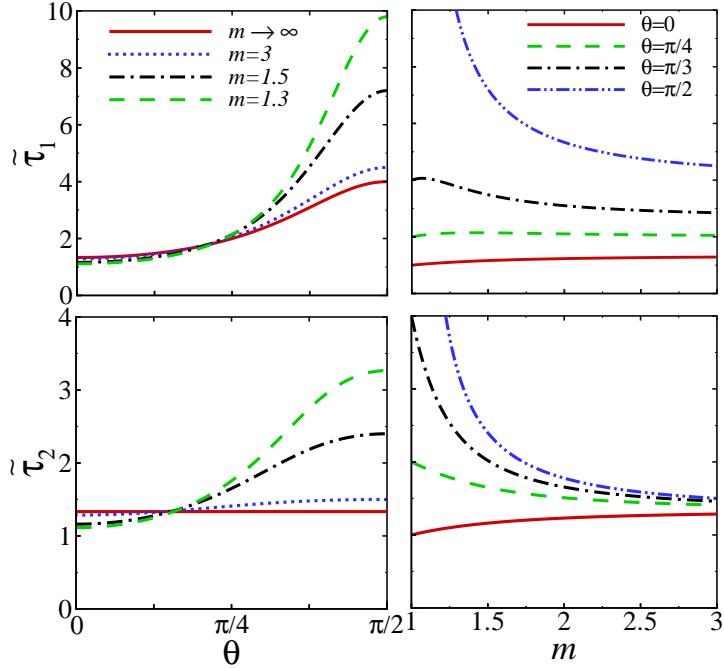}
\caption{(Color online)
Effective relaxation times ${\tilde \tau}_1$ and ${\tilde \tau}_2$ of the electrons at the Fermi surface, \textit{i.e.}, $\varepsilon=\mu$, in terms of $\theta$ for different values of $m$ (left) and in terms of $m$ for different values of $\theta$ (right). 
\label{relaxt1}}
\end{figure}

At $\theta=0$, when the spin of all impurities are perpendicular to the surface, the system is isotropic and we find $\sigma_{xx}=\sigma_{yy}$ for all $m$ values. At a fixed value of the chemical potential, the mass of electron is reduced by increasing $m$ and the band gap becomes narrower which leads to an increment in the conductivities. 
Increasing $\theta$, the back-scattering probability gradually diminishes, and the conductivity mounts up monotonically. 
We find that, at $\theta=\pi/2$, the conductivities have their maximum values, \textit{i.e.}, $\sigma_{xx}=\sigma_0$ and $\sigma_{yy}=\sigma_0/3$, independent of the value of $m$.
The reason for this constant conductivity could be understood through the effective mobility of the surface electrons $\mu_e$.
The mobility is a material-specific quantity which simultaneously includes features of both the scattering probability (through $\tau$) and the surface band structure (through $v_\kv$). Since our system is anisotropic, the effective mobilities of surface electrons are also modified as $\mu_e^1=\tilde{\tau}_1v_\kv^2$ and $\mu_e^2=\tilde{\tau}_2 v_\kv^2$, where $\tilde{\tau}_1=4\tau^c_{1,1}/\tau_\kv^0$ and $\tilde{\tau}_2=4\tau^s_{2,1}/\tau_\kv^0$ are the effective relaxation times ($\tau^c_{1,1}$ and $\tau^s_{2,1}$ are introduced in the Appendix). In order to study the behavior of these mobilities with respect to $\theta$ and $m$, we first investigate the behavior of the effective relaxation times $\tilde{\tau}_1$ and $\tilde{\tau}_2$.
We have plotted the effective relaxation times in terms of $\theta$ and $m$ in Fig.~\ref{relaxt1}. 
At small values of $\theta$, when the spins of impurities have large component along the $z$-axis, changing $m$ has no significant effect on the behavior of relaxation times (see, \textit{e.g.}, the curves for $\theta=0$ and $\theta=\pi/4$ in the right panel of Fig.~\ref{relaxt1}). However, by increasing $\theta$ they become more sensitive to the small values of $m$. At $\theta=\pi/2$, the effective relaxation times are $\tilde{\tau}_1(\varepsilon=\mu)=3\tilde{\tau}_2(\varepsilon=\mu)=4 /(1-m^{-2})$, and an enhancement of $m$ accompanies with a sharp reduction in the relaxation times. 
On the other side, the velocity $v_\kv=v_{\rm F}\sqrt{1-m^{-2}}$, 
is a signature of the band structure and does not depend on the direction of the spins of magnetic impurities.
Therefore the mobilities and consequently the conductivities $(\sigma=n e\mu_e)$  
become independent of $m$ at $\theta=\pi/2$.

In the insets of Figs.~\ref{cond-gapped1} and \ref{cond-gapped2} we have also plotted the anisotropic magneto-resistance in terms of the orientation of the spin of surface impurities $\theta$ and the inverse mass $m$. At a fixed value of the chemical potential, increasing the energy gap, \textit{i.e.}, decreasing $m$, a reduction is clearly seen in the AMR.
At the large values of $m$, the AMR is almost constant, but it experiences a sharp reduction at small $m$.

\section{Summary and Conclusion}\label{sec:summary}

In this paper, we have obtained expressions for the conductivity of gapless and gapped surface states of a magnetic topological insulator using semiclassical Boltzmann approach. Because of the anisotropic scattering of electrons from magnetic impurities at the surface of TI, the standard relaxation time approximation is not applicable anymore. In this paper we have used a modified scheme and obtained a closed form for the relaxation times of a magnetic topological insulator. Employing the closed forms of the relaxation times we have obtained the surface conductivities of the system. 

Finally, we would also like to comment briefly on extra contributions to the surface conductivities arising from
the anomalous velocity of the surface electrons, which causes a non-zero transverse conductivity in the system~\cite{sinitsyn_review}.
The anomalous velocity, ${\bf v}^{a}_{n}(\kv)= e \textbf{E}\times {\mathbf\Omega}_{n}(\kv)/\hbar$, depends on the topology of the band structure and is given in terms of the Berry curvature of the $n$-th Bloch band: $\mathbf{\Omega}_n=i\nabla_{\kv}\times\langle u_{n}({\kv})|\nabla_{\kv}| u_{n}(\kv)\rangle$,
where $u_n({\kv})$ is the cell periodic eigenstate of ${\cal H}_0(\kv)$. It is well known that in magnetic topological insulators the Berry curvature of the surface valence and conduction bands are different. In our system they are given in terms of the bulk magnetization $M$ as 
\begin{equation}
{\bf\Omega}_\pm(\kv)=\mp\frac{M\hbar^2 v^2_{F}}{2(\varepsilon_k^{\rm M})^3}\hat{e}_z~.
\end{equation}
The anomalous transverse conductivity at zero temperature is therefore obtained as
$\sigma_{xy}^a=-1/(2 m)$ (in the units of $e^2/h$)~\cite{sinitsyn_prb_2007}.
Note that the anomalous conductivity is independent of the impurity scattering and only depends on the chemical potential and the surface energy gap, and vanishes above  $T_c^{bulk}$, when a topological phase transition occurs by gap closing. 
Moreover, the side-jump effect, which arises from the spin-orbit modification to the position operator would also give a similar contribution to the transverse conductivity~\cite{amir_inpreparation}. 
The anomalous and side-jump contributions to the conductivity are of the order of $e^2/h$, while the extrinsic longitudinal conductivities reported here are normally several orders of magnitude larger (typically, $~ 10^2 - 10^4 ~e^2/h$).
Therefore, the transverse resistivity $\rho_{xy}\approx -\sigma_{xy}/(\sigma_{xx}\sigma_{yy})$ would be negligible in compare with the longitudinal ones, $\rho_{ii}\approx 1/\sigma_{ii}$.

\ack
J. A. gratefully acknowledges the hospitality of ICTP where part of this work was completed.
\appendix

\section{Calculation of the relaxation times} \label{sec:appendix}
In this section we explain the details of obtaining the generalized relaxation times $\tau_i$ ($i=1,2$) introduced in Eq.(\ref{rt-aniso}), employing their Fourier expansions as
\be\label{tau_fourier}
\tau_{i}(\phi_{\kv})= \sum_{n=1}^{\infty}[{\tau}^c_{i,n}\cos(n\phi_\kv)+{\tau}^s_{i,n}\sin(n\phi_{\kv})]~,
\ee
where ${\tau}^{c(s)}_{i,n}$ are the Fourier coefficients which are evidently independent of $\phi_\kv$. 
Now, substituting~(\ref{tau_fourier}) in Eq.~(\ref{rt-aniso}), and using the appropriate transition probability $W_{\kv,\kv'}$, one can find a closed forms for the generalized relaxation times. 
Note that, the conservation of particle number forbids $n=0$ terms in the Fourier expansions of $\tau_i$.

Below, we first explain the results for the gapless surface states. In this case one simply finds 
${\bar w}(\phi_\kv)=2/\tau_\kv^0$, with $\tau_\kv^0=4\hbar^3 v^2_{{\rm F}}/(n_{{\rm imp}} J^2 S^2 \varepsilon_k)$ as defined in the main text, is independent of
the direction of both the incident electron and the spin of impurity. 
Now, replacing~(\ref{tau_fourier}) in Eq.~(\ref{rt-aniso}), we find
\be
{ \tau}^c_{1,1}=\frac{\tau_\kv^0}{2+\cos 2 \theta}~,~~
{ \tau}^s_{2,1}=\frac{\tau_\kv^0}{3}~,
\ee
and all other coefficients are identically zero. Therefore we simply obtain 
\be
\tau_1(\phi_\kv)=\frac{ \tau_\kv^0 \cos\phi_\kv}{2+\cos 2 \theta}~,~~
\tau_2(\phi_\kv)=\frac{\tau_\kv^0 \sin\phi_\kv}{3} ~.
\ee
In the case of the gapped surface states the expressions turn out to become quite complicated.
Using Eq.~(\ref{T-gapped}), the averaged transition probability ${\bar w}(\phi_\kv)$ reads
\be\label{w_bar}
{\bar w}(\phi_\kv)=\frac{2}{\tau_\kv^0}(w^0+w^c \cos\phi_\kv)~,
\ee
where
\bea
\nonumber w^0&=& 1+\gamma_k^2\cos2\theta~,\\
w^c&=&\gamma_k(1-\gamma_k^2)^{1/2}\sin2\theta~.
\eea
Moreover, using the Fourier expansion of $\tau_i(\phi)$, it is straightforward to show that
\be\label{wt_bar}
\sum_{\kv'} W_{\kv,\kv'}\tau_i(\phi_{\kv'})=\alpha_i^0+\alpha^c_i\cos\phi_\kv+\alpha^s_i\sin\phi_\kv~,
\ee
where the coefficients are defined as
\bea
\alpha^0_i&=&  \frac{{ \tau}^c_{i,1}}{\tau_\kv^0} \gamma_k(1-\gamma_k^2)^{1/2}\sin2\theta~,\\
\alpha^c_i&=&-\frac{{ \tau}^c_{i,1}}{\tau_\kv^0} (1-\gamma_k^2)\cos2\theta~,\\
\alpha^s_i&=&-\frac{{ \tau}^s_{i,1}}{\tau_\kv^0} (1-\gamma_k^2)~.
\eea
Now, using expressions~(\ref{w_bar}) and~(\ref{wt_bar}) in Eq.~(\ref{rt-aniso}), we find
\bea
\tau_1(\phi_{\kv})  &=&\frac{\alpha_1^0+(1+\alpha_1^c)\cos\phi_\kv+\alpha_1^s\sin\phi_\kv}{2(w^0+w^c \cos\phi_\kv) }~\tau_\kv^0~,\label{tau1_closed}\\
\tau_2(\phi_{\kv})  &=&\frac{\alpha_2^0+\alpha_2^c\cos\phi_\kv+(1+\alpha_2^s)\sin\phi_\kv}{2(w^0+w^c \cos\phi_\kv) }~\tau_\kv^0~.\label{tau2_closed} 
\eea
Note that the general closed forms~(\ref{tau1_closed}) and (\ref{tau2_closed}) of the relaxation times could be determined only in terms of the coefficients ${ \tau}^c_{i,1}$ and ${ \tau}^s_{i,1}$. These coefficients could be obtained using
\bea
{ \tau}^c_{i,1}&=&\frac{1}{\pi}\int_0^{2\pi}\mathrm{d}\phi_\kv \cos\phi_\kv \tau_i(\phi_\kv)~,\\
{ \tau}^s_{i,1}&=&\frac{1}{\pi}\int_0^{2\pi}\mathrm{d}\phi_\kv \sin\phi_\kv \tau_i(\phi_\kv)~,
\eea
which leads to
\bea
{\tau}^c_{1,1}&=& \frac{\tau_\kv^0}{1+\Gamma+\left(1+\Gamma\gamma_k^2\right)\cos2\theta}~,\\
{ \tau}^s_{2,1}&=& \nonumber\frac{\tau_\kv^0}{1-\gamma_k^2+
(1+\Gamma)w^0} ~,\\
\eea
with $\Gamma=\sqrt{1-\left(w^c/w^0\right)^2}$, and ${ \tau}^s_{1,1}={ \tau}^c_{2,1}=0$.
Note that the longitudinal conductivities in general depend only on ${ \tau}^c_{1,1}$ and ${ \tau}^s_{2,1}$, and 
the transverse ones depend only on ${ \tau}^s_{1,1}$ and ${ \tau}^c_{2,1}$ and the general closed forms of the relaxation times are not important. 


\end{document}